\documentclass[%
 aip,
 amsmath,amssymb,
reprint,%
]{revtex4-1}

\usepackage{graphicx}% Include figure files
\usepackage{dcolumn}% Align table columns on decimal point
\usepackage{bm}% bold math
\usepackage[utf8]{inputenc}
\usepackage[T1]{fontenc}
\usepackage{mathptmx}

\usepackage{chemformula} % Formula subscripts using \ch{}
\usepackage[T1]{fontenc} % Use modern font encodings
\usepackage{graphics}% Include figure files
\usepackage{bm}% bold math
\usepackage{xfrac}
\usepackage{mathrsfs}
\usepackage{float}
\usepackage[normalem]{ulem}
\usepackage[bbgreekl]{mathbbol}
\usepackage[utf8]{inputenc}
\usepackage{amsmath} 
\usepackage{adjustbox}
\usepackage{etoolbox}
\usepackage{comment}
\usepackage{amsmath}
\usepackage{amssymb}
\usepackage{hyperref}
\usepackage{tikz}
\usetikzlibrary{matrix}

\DeclareMathOperator*{\SumInt}{%
\mathchoice%
  {\ooalign{$\displaystyle\sum$\cr\hidewidth$\displaystyle\int$\hidewidth\cr}}
  {\ooalign{\raisebox{.14\height}{\scalebox{.75}{$\displaystyle\sum$}}\cr\hidewidth{\scalebox{.75}{$\displaystyle\int$}}\hidewidth\cr}}
  {\ooalign{\raisebox{.2\height}{\scalebox{.6}{$\scriptstyle\sum$}}\cr$\scriptstyle\int$\cr}}
  {\ooalign{\raisebox{.2\height}{\scalebox{.6}{$\scriptstyle\sum$}}\cr$\scriptstyle\int$\cr}}}
\newcommand{\EP}[0]{\mathrm{EP}}
\newcommand{\figref}[1]{Fig.~\ref{fig:#1}}
\newcommand{\figrefbegin}[1]{Figure~\ref{fig:#1}}

\renewcommand{\eqref}[1]{Eq.~(\ref{eq:#1})}
\newcommand{\eqrefbegin}[1]{Equation~\ref{eq:#1}}

\newcommand{\mat}[1]{\mathbb{#1}}
\newcommand{\BRA}[1]{({#1}|}
\newcommand{\KET}[1]{|{#1})}
\newcommand{\ket}[1]{|{#1}\rangle}
\newcommand{\bra}[1]{\langle{#1}|}
\newcommand{\bracket}[2]{\langle{#1}|{#2}\rangle}
\newcommand{\BBRA}[1]{(({#1}|}
\newcommand{\KKET}[1]{|{#1}))}
\newcommand{\BBRAKKET}[2]{(({#1}|{#2}))}
\newcommand{\BRACKET}[2]{({#1}|{#2})}
\newcommand{\BRAMKET}[3]{({#1}|{#2}|{#3})}

\newcommand{\citeasnoun}[1]{Ref.~\citenum{#1}}
\newcommand{\citeasnouns}[1]{Refs.~\citenum{#1}}

\begin{document}

\preprint{AIP/123-QED}

\title[] {Ab-initio  Theory  of  Photoionization via Resonances}
% Force line breaks with \\

\author{Adi Pick}
\altaffiliation[Also at ]{Faculty of Electrical Engineering, Technion-Israel Institute of Technology, Haifa, Israel.}
\email{pick.adi@gmail.com}
\affiliation{Faculty of Chemistry, Technion-Israel Institute of Technology, Haifa, Israel.}
\author{Petra Ruth Kapr{\'a}lov{\'a}-{\v{Z}}{\v{d}}{\'a}nsk{\'a}}
\affiliation{Department of Radiation and Chemical Physics, Institute of Physics, Academy of Sciences of the Czech Republic, \\Na Slovance 2, 182 21 Prague 8, Czech Republic}
\author{Nimrod Moiseyev}
\altaffiliation[Also at ]{Faculty of Physics, Technion-Israel Institute of Technology, Haifa, Israel.}
\affiliation{Faculty of Chemistry, Technion-Israel Institute of Technology, Haifa, Israel.}

\date{\today}% It is always \today, today,

\begin{abstract}
 We present an \emph{ab-initio} approach for computing the photoionization spectrum  near autoionization   resonances  in  multi-electron systems.  While traditional   (Hermitian) theories   typically require computing  the continuum states, which   are difficult to obtain with high accuracy, our non-Hermitian approach requires only  discrete bound and metastable states, which can be accurately computed with   available quantum chemistry tools.     We derive a simple  formula for the  absorption lineshape near Fano resonances, which relates the asymmetry  of the spectral peaks  to the phase of the complex transition dipole moment. Additionally, we present a formula for the ionization spectrum of  laser-driven targets and    relate  the ``Autler-Townes'' splitting of spectral lines to the existence of exceptional points in the Hamiltonian.  We apply our formulas to compute the autoionization spectrum of helium, but our theory is also applicable  for non-trivial     multi-electron    atoms  and molecules.
\end{abstract}

\maketitle

%---------------------------------------------
%\section{Introduction}
%---------------------------------------------

We present an \emph{ab-initio} approach for computing the photoionization spectrum  near autoionization (AI) resonances  in  multi-electron systems.  Recent   developments in attosecond-laser technology  enable probing and controlling photoionization processes, and lead to  a renewed interest in ionization  and  related phenomena, such as high-harmonic generation and strong-field electronic dynamics~\cite{krausz2009attosecond,itatani2004tomographic,paul2001observation,klunder2011probing,azoury2017self}.  These experimental capabilities  call for \emph{ab-initio} theories, which  can  relate the electronic structure of the sampled medium  to the measured ionization spectrum. However, most  existing theories require the calculation of the continuum states~\cite{feshbach1962unified,friedrich1985interfering} (above the ionization threshold), which are difficult to obtain with high accuracy with traditional  methods~\cite{averbukh2005ab,goetz2017theoretical}.  In this work, we use  non-Hermitian  quantum mechanics (NHQM)~\cite{Moiseyev2011} in order to   avoid the need of computing   continuum states.  Our theory produces a simple formula for the ``Fano asymmetry parameter'' [\eqref{our-fano}], which expresses the asymmetry of the peaks  in the ionization spectrum near AI resonances~\cite{fano1961effects}, 
and a formula for the photoionization  spectrum of laser-driven   systems~[\eqref{floquet-absorption-formula}]. We relate the  Autler-Townes~\cite{autler1955stark} splitting of  ionization  spectral peaks      to the  existence of  exceptional points~\cite{kato2013perturbation} (EPs)---special degenerate resonances where multiple   AI states  share the same energy and wavefunction. We  demonstrate the predictions of our theory for helium using \emph{ab-initio} electronic-structure  data   from~\citeasnoun{kapralova2013excitation,kapralova2013gaussian}.
By using  advanced   non-Hermitian quantum-chemistry algorithms~\cite{zuev2014complex,landau2016atomic},  our theory can also be applied for larger atoms and molecules.

\begin{figure}[t]
\centering
    \includegraphics[scale=0.7]{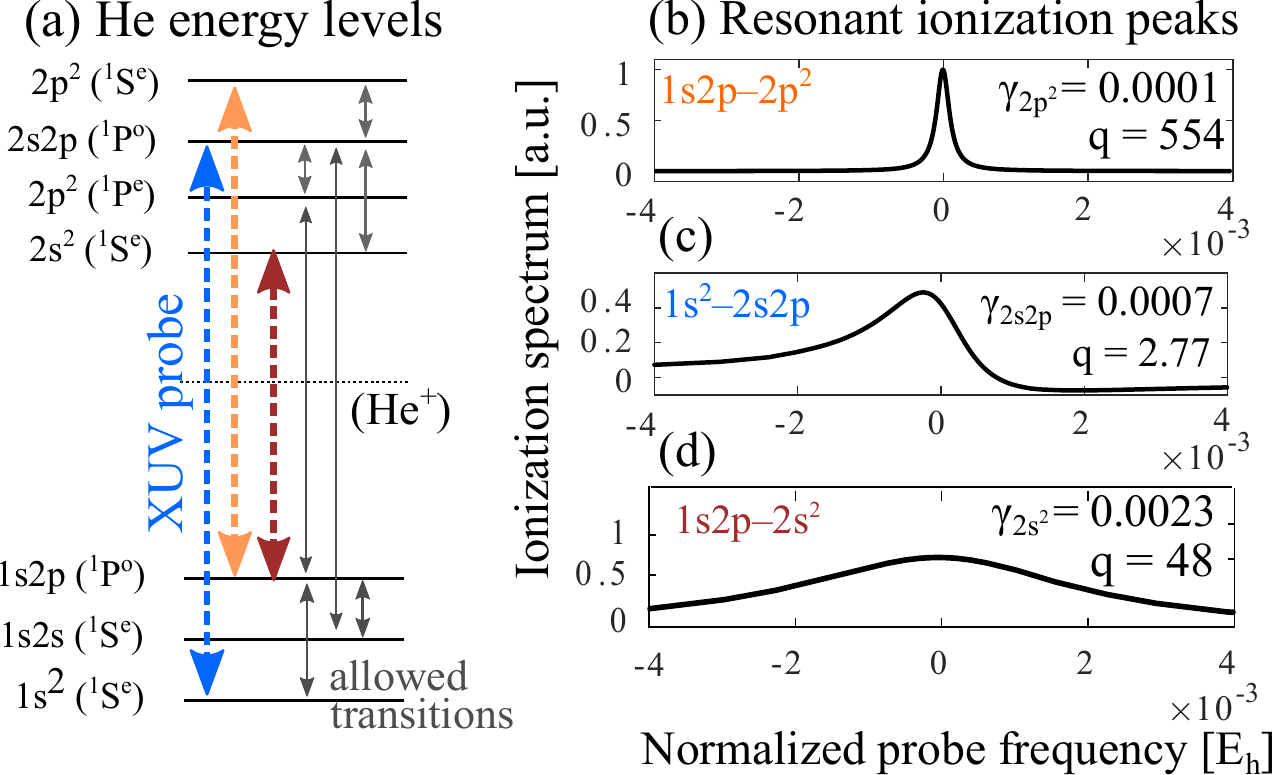}
  \caption{    (a)  Orbitals of  parahelium (labeled by   spectroscopic notation~\cite{levine1991quantum}), including three bound states and four autoionization resonances (above the ground-state energy of He$^+$). Thick dashed arrows mark the probed   transitions, while narrow arrows mark additional dipole-allowed transitions. 
   (b--d) Autoionization spectrum [evaluated  using~\eqref{ff-formula} 
   using data from~\citeasnoun{kapralova2013gaussian}, summarized in Table S1 in the SI] near  the three probed transitions from (a) and the associated   $q$ factors [evaluated using \eqref{our-fano}].
   The $x$-axis is the  difference   between the frequency of the field and the probed atomic transition. [e.g., in (b), ``zero frequency'' implies that the probe is resonant with the $1s2p\leftrightarrow2p^2$ transition.]  }
  \label{fig:helium-figure}
\end{figure}

 In NHQM, the time-independent Schr\"{o}dinger  equation is solved with outgoing boundary conditions and the resulting energy spectrum is discrete,  containing real-energy bound states and complex-energy metastable   states~\cite{Moiseyev2011}. This situation is very different from traditional Hermitian quantum mechanics (HQM), where  metastable (autoionizing) states are described as    real-energy bound states embedded in a    continuum of free  states~\cite{fano1961effects}. Moreover, in HQM, the transition dipole moment is real, while in NHQM, it is generally complex.    We show  that {the phase of the complex transition dipole moment   has  physical significance:} it determines the asymmetry  of  the  absorption peaks near  AI states (see \figref{helium-figure}). Our derivation of the Fano asymmetry parameter is inspired by the recent work of Fukuta~\emph{et al.}~\cite{fukuta2017fano}, which used a non-Hermitian approach to compute the Fano factor, although~\citeasnoun{fukuta2017fano} introduced an artificial model for the continuum and did not relate the lineshape to the complex transition dipole moment.

\begin{figure}[h]
  \begin{center}
  \includegraphics[scale=0.6]{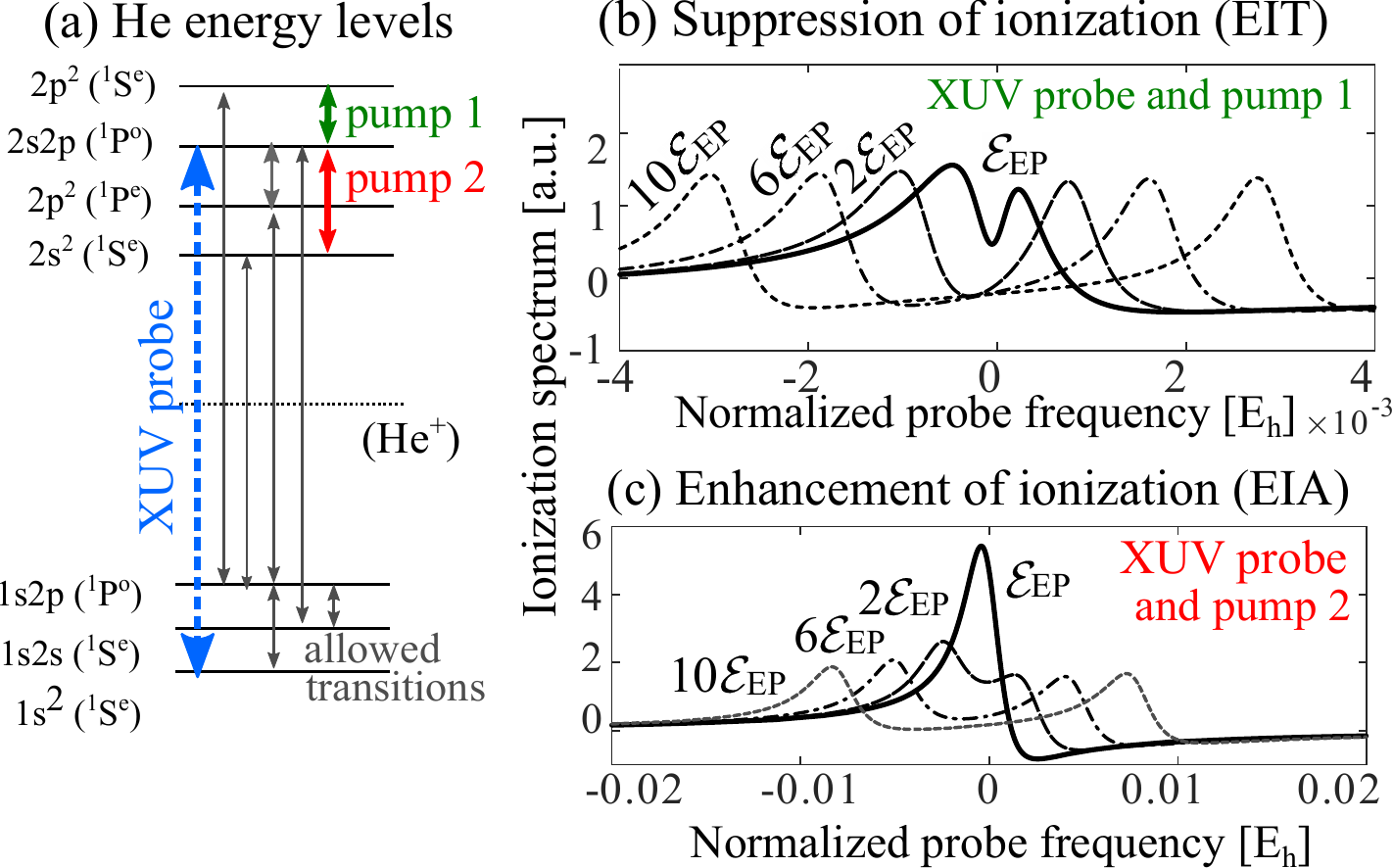}
  \end{center}
  \caption{(a) Orbitals  of parahelium (as in \figref{helium-figure}).
   Thick dashed (solid) arrows mark the probed (pumped) transitions, while  narrow arrows mark other  allowed transitions.
(b) Pump 1 couples  the probed AI state  [$2s2p (^1P^o)$] to the state  $2p^2({^1}S^e)$, whose lifetime is longer than $2s2p$.
The pump produces  a dip in the  ionization spectrum  when the probe is resonant with the probed transition,  similar to EIT. 
(The $x$-axis is  the probe's detuning  from resonance, as in~\figref{helium-figure}.)
 Four pump amplitudes are shown: $\mathcal{E}/\mathcal{E}_\mathrm{EP}= 1, 2, 6, 10$, while the pump intensity is  $\omega_2^\mathrm{EP}$ (see text).   Helium parameters are summarized in Table S1.
  (c) Pump 2 couples the probed AI state  to the state $2s^2 (^1S^e)$, whose  lifetime is shorter than $2s2p$. 
 The resulting ionization spectrum   shows  enhancement  of ionization on resonance (similar to EIA).}
  \label{fig:EPIT-EPIO}
\end{figure}

As an application of our approach, we study the suppression and enhancement of photoionization in the presence of  an external driving laser [\figref{EPIT-EPIO}]---two effects which  are closely related to electromagnetically induced transparency (EIT)~\cite{boller1991observation,harris1990nonlinear,harris1997electromagnetically,fleischhauer2000dark,fleischhauer2005electromagnetically} and absorption (EIA)~\cite{lezama1999electromagnetically,goren2003electromagnetically,zhang2015electromagnetically}. Laser-induced suppression of photoionization     was first measured  in  magnesium~\cite{karapanagioti1995observation,karapanagioti1996effects}, and later demonstrated  in several other systems~\cite{halfmann1998population,gao2000electromagnetically}. 
These  experiments    were  modeled using the   Feshbach formalism~\cite{bachau1986theory,karapanagioti1996effects}, which requires the computation of the continuum states  and can be avoided by our  approach. 
 Motivated by   the huge impact of EIT  in  optics and atomic physics~\cite{harris1990nonlinear,harris1997electromagnetically,fleischhauer2000dark,fleischhauer2005electromagnetically}, we believe that  the ability to accurately compute the analogous effects for AI states    will open new routes for controlling and understanding  photoionization and  related phenomena (e.g., photoassociation and photodetachment).

We  begin by reviewing the  traditional theory of photoionization~\cite{fano1961effects}. 
In order to obtain the photoionization spectrum, one needs to compute   the rate  at which an absorbed photon excites an atom (or molecule) into an AI state.  
The transition rate induced by a linearly polarized electromagnetic field (with frequency  $\omega$, amplitude $\mathcal{E}$, and polarization axis $\hat{x}$) can be computed using  Fermi's golden rule~\cite{yan1986eigenstate}
\begin{gather}
S(\omega) = \frac{|\mathcal{E}|^2}{\hbar}\SumInt_f 
\left|\langle{\phi_i}|x|{\phi_f}\rangle\right|^2\delta(\hbar\omega-\varepsilon_f+\varepsilon_i).
\label{eq:FGR}
\end{gather}
Here,  $\varepsilon_{i,f}$ and $\ket{\phi_{i,f}}$ are the initial- and final-state energies and wavefunctions respectively, while  $\SumInt_f$  denotes summation over  bound      and integration over   continuum final   states. 
The eigenfunctions and energies are  obtained by diagonalizing the Hamiltonian (e.g., by  using standard quantum-chemistry tools~\cite{schmidt1993general}). 
Since the continuum states have delocalized wavefunctions, it is challenging to compute them with the same level of accuracy as   the bound states~\cite{averbukh2005ab,goetz2017theoretical}.

  This difficulty can be   circumvented by using    the  Green's function eigenstate-free method (developed in~\citeasnouns{mukamel1999principles,yan1986eigenstate}).  
In this approach, one first decomposes the Hamiltonian of the system into the ``initial- and excited-state   Hamiltonians,'' defined as 
$\mat{H}_i \equiv \varepsilon_i\ket{\phi_i}\bra{\phi_i}$ and  $\mat{H}_\mathrm{exc} \equiv \SumInt_f \varepsilon_f\ket{\phi_f}\bra{\phi_f}$   respectively. 
The key idea of this approach is to realize that the summation in \eqref{FGR} can be avoided by exploiting the ``excited-states' Green's function,'' which is the impulse response of the operator $\mat{H}_\mathrm{exc}$, defined as~\cite{trefethen1997numerical} 
\begin{gather}
\mat{G}_\mathrm{exc}(\varepsilon) \equiv (\mat{H}_\mathrm{exc}-\varepsilon)^{-1} =
 \SumInt_f 
 \frac{|{\phi_f}\rangle\langle{\phi_f}|}{\bracket{\phi_f}{\phi_f}}
\frac{1}{\varepsilon - \varepsilon_f}.
\label{eq:G-in-HQM}
\end{gather}
In  the second equality, we employ  the  normal-mode expansion of the Green's function~\cite{Arfken2006} . 
  Using a mathematical identity for the $\delta$-function~\cite{shibatani1968antiresonance}, one can  rewrite \eqref{FGR} as
\begin{align}
S(\omega) = 
\SumInt_f \langle{\phi_i}|x|{\phi_f}\rangle\left(
\tfrac{1}{\hbar\pi}\mathrm{Im}\!\lim_{s\rightarrow0^+}
\tfrac{|\mathcal{E}|^2}{\hbar\omega - is-\varepsilon_f+\varepsilon_i}\right)\langle{\phi_f}|x|{\phi_i}\rangle,
\label{eq:FGR-step1}
\end{align}
where the modes are normalized to one (i.e., $\bracket{\phi_f}{\phi_f} = 1$). 
Then, by substituting  \eqref{G-in-HQM}  into \eqref{FGR-step1}, one obtains  
\begin{gather}
S(\omega) = \frac{|\mathcal{E}|^2}{\hbar\pi}\mathrm{Im}\langle{\phi_i}|x\,\,\mat{G}_\mathrm{exc}\,x|{\phi_i}\rangle,
\label{eq:Green-absorption} 
\end{gather}
where $\mat{G}_\mathrm{exc}(\varepsilon)$ is evaluated at $\varepsilon = \hbar\omega +\varepsilon_i$.

Although \eqref{Green-absorption} does not contain summation over continuum states [and is, therefore, more efficient than  \eqref{FGR} in many cases], obtaining  the excited-state  Green's function  of multi-electron systems  requires significant computation power.  However,  the  Green's function formulation  naturally   extends to NHQM, which offers a huge computational advantage.
Under the conditions stated below, one can  replace   the Hermitan \emph{normal-mode} expansion of $G_\mathrm{exc}$ [\eqref{G-in-HQM}] with the non-Hermitian  \emph{quasi-normal mode} expansion~\cite{Moiseyev2011,trefethen2005spectra}:
\begin{equation}
G_\mathrm{exc}(\varepsilon) = \sum_f \frac{\KET{\phi^R_f}\BRA{\phi_f^L}}
{\BRACKET{\phi_f^L}{\phi_f^R}}
\frac{1}{\varepsilon - \varepsilon_f}.
\label{eq:modal-expansion-G}
\end{equation}
Here, $\KET{\phi_f^{R}}$ and $\varepsilon_f$   are the  discrete  eigenvectors and eigenvalues of the (non-Hermitian) time-independent Schr\"{o}dinger   equation---solved   with outgoing boundary conditions---and   $\KET{\phi_f^{L}}$ are the eigenvectors of the transposed Schr\"{o}dinger equation.  Round brackets denote the ``unconjugated  norm:''  $\BRACKET{\phi}{\psi}\equiv\int\!dx\,\phi\psi$~\cite{Moiseyev2011}, which generalizes the traditional Dirac ``conjugated norm,'' $\langle\phi|\psi\rangle\equiv\int\!dx\,\phi^*\psi$~\cite{Griffiths2005} from HQM.  Since the  energy spectrum in NHQM is discrete, we replace  $\SumInt$ by $\sum$.

\eqrefbegin{modal-expansion-G} is valid assuming that   
($i$)  the Green's function is evaluated   near the resonant energies (i.e., when $\varepsilon\approx\mathrm{Re}[\varepsilon_f]$), 
($ii$) the excited AI states, ${\phi^R_f}(x)$, are evaluated near the   interaction region~\cite{lee1999dyadic,leung1994completeness} (i.e., not far from the atomic core, $|x|\approx a_0$, where $a_0$ is the Bohr radius), and 
($iii$) when the  energy spectrum does not contain EPs. 
Condition ($i$) is satisfied (in all cases of interest)  since we apply \eqref{modal-expansion-G} only to study  resonant absorption.  
Condition ($ii$) is fulfilled because the  absorption formula [\eqref{Green-absorption}]    depends  only on the    overlap  integral  between extended AI states, ${\phi^R_f}(x)$,  and localized bound states, ${\phi^R_i}(x)$.
Although the eigenvectors of the non-Hermitian Hamiltonian, ${\phi^R_f}(x)$, are typically  poor approximations for  the  ``physical'' metastable-state wavefunctions far from the interaction region (where they have  diverging tails~\cite{Moiseyev2011}), their unphysical values  at large distances  do not contribute to the absorption spectrum since the bound-state wavefunctions vanish at large distances.
Last, when the energy spectrum contains EPs,~\eqref{modal-expansion-G} breaks down because the quasi-normal modes, ${\phi^R_f}(x)$, do not form a complete basis of the Hilbert space~\cite{Hanson2003,Hernandez2000,Hernandez2003,pick2017general}. At an EP, 
the ``unconjugated norm'' of the degenerate state, $\BRACKET{\phi_\mathrm{EP}^L}{\phi_\mathrm{EP}^R}$,   vanishes  and the  associated term in  \eqref{modal-expansion-G}  blows up. One can obtain a corrected formula for $G_\mathrm{exc}$ at the EP by considering \eqref{modal-expansion-G} \emph{near the EP} and carefully taking the limit of approaching the EP.
In this limit, the two terms in the sum that are associated with the nearly degenerate  poles dominate the impulse response due to their infinitesimal denominators. However,  one finds that two terms diverge with opposite signs, while  their sum remains finite.  This point was previously realized on pure mathematical grounds~\cite{trefethen2005spectra}, and later explained  in the context of  electromagnetic modes~\cite{Hanson2003,Hernandez2000,Hernandez2003,pick2017general}. We sketch the derivation of  the corrected formula for $G_\mathrm{exc}$  in the context of AI states  in the supporting information  (SI) appendix.

By substituting \eqref{modal-expansion-G} into \eqref{Green-absorption}, we obtain the  NHQM    absorption spectrum formula:
\begin{gather}
S(\omega) = 
\frac{\left|\mathcal{E}\right|^2 }{\hbar\pi}\mathrm{Im}
\sum_f
\frac{ \BRA{\phi_i^L}x \KET{\phi_f^R}
\BRA{\phi_f^L}x \KET{\phi_i^R}}
{\BRACKET{\phi_f^L}{\phi_f^R}(\hbar\omega-\varepsilon_f+\varepsilon_i)}.
\label{eq:ff-formula}
\end{gather}
This formula applies to  cases where the initial state is bound and, therefore, one  can  replace  Dirac bracket states, $|\phi_i\rangle$ and  $\langle\phi_i|$,   with  unconjugated bracket states $\KET{\phi_i^R}$ and $\BRA{\phi_i^L}$ (since the left eigenvector of a Hermitian Hamiltonian is equal to the conjugated right eigenvector of   the same eigenvalue~\footnote{Right and left eigenvectors satisfy the equations $\mat{H}\psi_i^R = \lambda_i\psi_i^R$ and $\mat{H}^T\psi_i^L = \lambda_i\psi_i^L$ respectively, where $T$ denotes matrix transposition.  When  $\mat{H}$ is Hermitian, one obtains 
$\mat{H}^*\psi_i^L = \lambda_i\psi_i^L$. Also, the eigenvalues of Hermitian operators are real.  By  complex-conjugating  both sides of the last equation, one obtains $\mat{H}(\psi_i^L)^* = \lambda_i(\psi_i^L)^*$, which proves that $(\psi_i^L)^*\propto\psi_i^R$. }). However,  when  the initial state is metastable,  one needs to keep the conjugated bracket for the initial state. Our derivation of \eqref{ff-formula} is similar in spirit to  the recent work    of Fukuta \emph{et al.}~\cite{fukuta2017fano}, which used NHQM to compute the absorption spectrum. However, the latter work  introduced an artificial model for   the continuum (which limits its generality)   and did not relate the  spectrum directly to the complex transition dipole moment.

{\eqrefbegin{ff-formula}  applies both  for autoionizing atoms and molecules. In either case, the AI rate depends on the complex transition moment between  initial and final  states of the system. While in atoms, ionizing transitions involve two different electronic states, in molecules (where the  wavefunction depends both on  electronic and  nuclear coordinates), there are, generally speaking,  two possibilities: 
($i$) Either the initial and final electronic states are different but the nuclear rovibrational state is the same (as in penning ionization~\cite{penning1927ionisation,miller1970theory,henson2012observation,bhattacharya2017polyatomic} or interatomic Coulombic decay~\cite{cederbaum1997giant,santra2000interatomic,averbukh2004mechanism}) or 
($ii$)  the initial and final electronic states are the same but the nuclear rovibrational state is different (as in dipole-bound anions~\cite{edwards2012vibrational} or intermolecular vibrational-energy transfer~\cite{cederbaum2018ultrafast}).
In case ($i$) (which involves transitions between  different electronic state), ionization can occur provided that the   energy of the excited electronic state  exceeds  the ground-state energy of the ion (i.e., the ionization threshold); under this condition, the excited state is degenerate with the continuum of free electronic  states.  Such processes are accurately described within the Born--Oppenheimer (BO) approximation, which amounts to assuming that the motion of atomic nuclei and electrons in a molecule can be separated~\cite{levine1991quantum}.  In  case ($ii$) (which  involves different nuclear rovibrational states),  ionization  happens due to non-adiabatic coupling terms (i.e.,  corrections beyond  the BO approximation), which provide coupling between the electronic states of the molecule in different nuclear configuration. 
In order for AI to occur in such cases,  the rovibrational  energy must exceed the binding energy of the electron to the molecule. 
While the latter situation has been treated  in~\citeasnoun{edwards2012vibrational}, our new formula [\eqref{ff-formula}] applies to  all cases. Note that~\citeasnoun{edwards2012vibrational} uses Hermitian scattering theory, while  our approach  uses  NHQM, which makes the spectrum discrete and, therefore, produces a computationally efficient formula. }

Our non-Hermitian   formula [\eqref{ff-formula}] provides a simple formula  for the ``Fano asymmetry parameter,'' which expresses the asymmetry of the peaks  in the ionization spectrum near AI resonances~\cite{fano1961effects}. According to traditional absorption theory~\cite{fano1961effects,shibatani1968antiresonance},  the spectrum near   an isolated  resonance with  frequency $\Omega$ and lifetime  $1/\gamma$ can be written as 
\begin{equation}
S_\mathrm{F}(\omega) = S_0(\omega)\,\frac{(\omega-\Omega + \tfrac{\gamma }{2} q)^2}{(\omega-\Omega)^2+(\tfrac{\gamma}{2})^2},
\label{eq:fano-original}
\end{equation}
where  $S_0(\omega)$ is the background absorption due to  continuum states and the remaining expression  is the resonant   peak. The parameter  $q$ determines the asymmetry of the resonant peak: in the limit of $q\rightarrow\infty$, the lineshape is  Lorentzian, while the limit of $q\simeq1$ produces  an  asymmetric lineshape. 
In traditional HQM, $q$ is found by computing overlap integrals involving   bound and continuum states~\cite{fano1961effects,shibatani1968antiresonance}. In our approach, however,  the $q$ factor depends  on a single   term in   \eqref{ff-formula} (for which $\mathrm{Re}[\varepsilon_f - \varepsilon_i]\approx\hbar\Omega$). By taking the ratio of the symmetric and anti-symmetric parts of  that  term (similar to~\citeasnoun{fukuta2017fano}), we obtain the compact expression
\begin{equation}
q =
\left|
 \frac{\mathrm{Re}\,\mu_{if}^2}{\mathrm{Im}\,\mu_{if}^2}
\right|
\left[
1 \pm
\sqrt{1+
\left( \tfrac{\mathrm{Im}\,\mu_{if}^2}{\mathrm{Re}\,\mu_{if}^2}\right)^{2}
}
\right],
\label{eq:our-fano}
\end{equation}
where   we introduced a shorthand notation  for the squared transition dipole moment, $\mu_{if}^2 \equiv \BRA{\phi_i^L}x \KET{\phi_f^R}
\BRA{\phi_f^L}x \KET{\phi_i^R}$. 
The sign of $q$  indicates whether the  absorption peak is blue or red shifted, and  is determined by  the sign of $\mathrm{Im}\,\mu_{ij}^2$.
More  details on the derivation are given in  the SI.  
{Our formula for $q$ generalizes  an earlier result from~\citeasnoun{edwards2012vibrational},  which  analyzed molecular photodetachment.   
The formulas agree qualitatively in the limit of large $q$, which corresponds to nearly Lorentzian AI peaks.
 While in our case, this limit is attained for nearly real resonant energies, $\varepsilon_f$ (which are associated with nearly real transition dipole moments, $\mu_{ij}$), in~\citeasnoun{edwards2012vibrational}, the ionization takes place due to non-adiabatic couplings and the limit of large $q$  emerges when the BO approximation is nearly accurate (i.e., when the non-adiabatic couplings are small).  A quantitative comparison of the formulas and an application of our formula for molecular photodetachment will be addressed in future work.}

%------------------------------------------------------
% FIGURE 1
%------------------------------------------------------
%------------------------------------------------------

As an example for  application of  \eqref{our-fano}, we compute the  absorption lineshape near AI states in parahelium (i.e., helium atoms  in which the spins of the two electrons are in the singlet state). The energy levels and dipole-allowed transitions are shown in \figref{helium-figure}(a). The states are labeled according to  the  approximate  Hartree--Fock orbitals. All   states below the ionization threshold    are bound, while all double-excitation states  are   metastable  AI states. We use  data from \citeasnouns{kapralova2013excitation,kapralova2013gaussian} for the energy levels, lifetimes, and complex transition dipole moments (summarized in Table S1 in the SI).   \figrefbegin{helium-figure}(b--d)  shows the ionization   spectrum  near  three transitions in the XUV range,  obtained by evaluating \eqref{ff-formula}. 
The $x$-axis denotes the frequency offset between the  field and the probed atomic transition.
The plots demonstrate that our new formula for the Fano  $q$ factor [\eqref{our-fano}]    predicts the asymmetry of the lines, while the width of the peak is set  by the imaginary part of the metastable-state energy.

 Next, we turn to study photoautoionization in  laser-driven atoms and molecules. Specifically, we consider cases where a ``pump  laser'' couples two (or more) AI states and a  ``probe laser''  drives transitions from the ground to the AI states. 
A time-periodic probed system (denoted below by $H_0$) has Floquet solutions  the form~\cite{dittrich1998quantum}:
\begin{equation}
\Psi_{\alpha,m}(x,t) = e^{-i\varepsilon_{\alpha,m} t/\hbar}\Phi_{\alpha,m}(x,t).
\label{eq:floquet-states}
\end{equation}
Here,  $\Phi_{\alpha,m}$ and $\varepsilon_{\alpha,m}$ are the eigenvectors and eigenvalues of the Floquet Hamiltonian, $\mathcal{H}\equiv H_0 - i\hbar\frac{\partial}{\partial t}$. The eigenenergies  are periodic in the frequency of the probe, $\omega_0$ (i.e., $\varepsilon_{\alpha m} = \varepsilon_{\alpha,0}+m\hbar \omega_0$) and the eigenvectors obey  
  $\Phi_{\alpha m}(t) = \Phi_{\alpha,0}(t)e^{i\omega_0 mt}$. The quantum number     $m$ is called   the ``Floquet channel.'' In  the SI, we use a generalized  Fermi-Floquet golden rule~\cite{bilitewski2015scattering} (which is valid for weak probe intensities) to derive a formula for  the absorption      spectrum of laser-driven systems. We obtain
\begin{gather}
S(\omega) = 
\frac{\left|\mathcal{E}\right|^2 }{\hbar\pi}\mathrm{Im}
\sum_{fm}
\frac{ \BBRA{\Phi_{i,0}^{L}}x \KKET{\Phi_{f,m}^{R}}
\BBRA{\Phi_{f,m}^{L}}x \KKET{\Phi_{i,0}^{R}}}
{\hbar\omega+m\hbar\omega_0-\varepsilon_{f,0}+\varepsilon_{i,0}}.
\label{eq:floquet-absorption-formula}
\end{gather}
Here  $\KET{\Phi_{i,0}}$  and   $\KET{\Phi_{f,m}}$ are the initial (bound) and final (metastable) Floquet states while    $\varepsilon_{i,0}$ and $\varepsilon_{f,0}$ are the corresponding quasienergies  in the zeroth Floquet channel. We use  double brackets  to denote spatial and temporal integration:  $\BBRAKKET{\Phi}{\Psi} \equiv \frac{1}{T}\int_0^T dt'\!\int_{-\infty}^{\infty}dx\,\, {\Phi(t',x)}{\Psi(t',x)}$.  In order to evaluate   \eqref{floquet-absorption-formula}, it is convenient to expand  the Floquet states,  $\KET{\Phi_{f,m}}$,    in the basis of  eigenvectors of the  field-free Hamiltonian.
% (obtained by subtracting from $H_0$ the term that is associated with the driving laser). 
  In the SI, we review this standard procedure~\cite{chu1977intense,chu1985recent} and present an  explicit expression for the spectrum in terms of  field-free eigenstates   [Eq.~(C22)].

%------------------------------------------------------
% FIGURE 2
%------------------------------------------------------
%----------------------------------------------------

Next, we  apply \eqref{floquet-absorption-formula}  to compute the autoionization spectrum of   laser-driven parahelium.  As shown in \figref{EPIT-EPIO}(a), we  consider  an XUV   probe, which drives the transition between the ground state  [$1s^2\,(^1S)$]  and the AI state $2s2p  (^1P)$ and a strong  NIR pump, which resonantly couples two AI states. We consider two cases: 
$(i)$ The pump couples   $2s2p \,(^1P)$ and   $2p^2\,(^1S)$ (green arrow) and 
$(ii)$ The pump couples $2s2p\,  (^1P)$ and $2s^2\,(^1S)$ and   (red arrow).  The difference between  the two cases     is that in the former, the probe couples  the ground state to  the broader AI state (out of the two coupled AI states), while in the latter, the probe couples the ground state to the narrower AI state. It turns out that these two situations lead to drastically different ionization spectra. In the former case [shown in \figref{EPIT-EPIO}(b)],  the effect of the pump is to suppress the ionization when  the probe-photon energy is resonant with the atomic transition. The narrow dip in the autoionization spectrum  is similar in spirit to the  transparency window in EIT. As shown in the early work of~\citeasnoun{karapanagioti1995observation},  the suppression  occurs due to coherent trapping of the atomic population  in the ground state.  In contrast, in the latter case [shown in \figref{EPIT-EPIO}(c)], the ionization   lineshape is Lorentzian for weak pump amplitudes [similar to electromagnetically induced absorption (EIA)]. As the pump exceeds a critical value (denoted by  $\mathcal{E}_0 = \mathcal{E}_{\mathrm{EP}}$), the peak splits into two non-overlapping dressed-state peaks. At the splitting point, the Hamiltonian has an EP, as discussed  below.
Panels (b--c) show the splitting of the ionization  peaks upon increasing the pump amplitude beyond the critical  point  (we show four pump values $\mathcal{E}_0/\mathcal{E}_{\mathrm{EP}} = 1,2,6,10$).  Our \emph{ab-initio}  calculation  takes into account all the dipole-allowed transitions in parahelium, although the atomic population predominantly occupies only three states in all  cases  under study.

%------------------------------------------------------
% FIGURE 3
%------------------------------------------------------
  \begin{figure}[t]
\centering
    \includegraphics[scale=0.6]{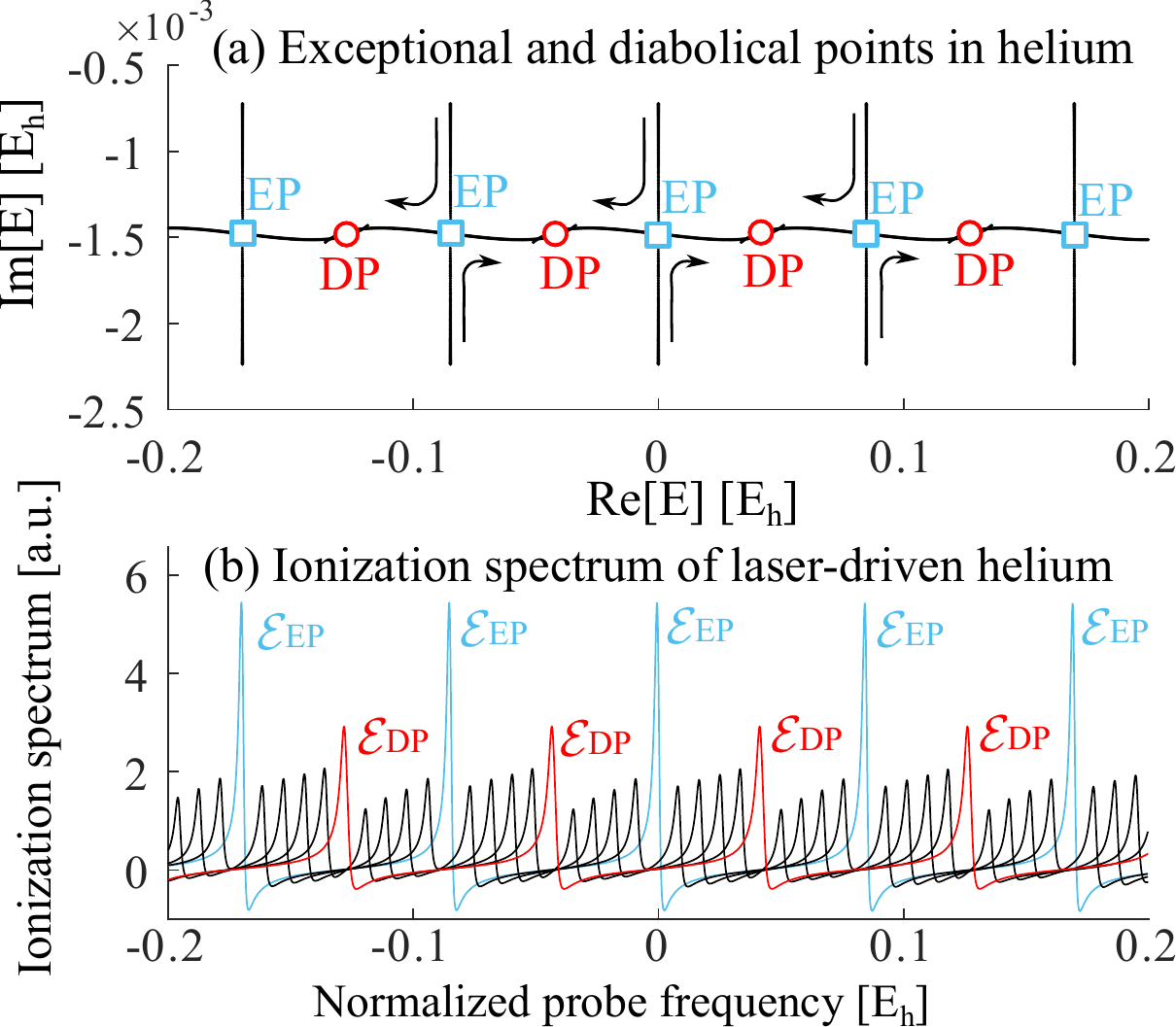}
  \caption{(a) Trajectories of   eigenvalues of the Floquet Hamiltonian [Eqs.~(C6-C7)]  upon varying the pump amplitude across the EP value while fixing the pump frequency at $\omega_2^\mathrm{EP}$ (see text). Helium parameters are saummarized in Table S1. Pairs of eigenvalues merge at  exceptional and diabolical points (EPs and DPs), marked by cyan triangles and red circles respectively. 
  (b) Relative absorption as a function of probe detuning for six pump amplitudes: $\mathcal{E}_0 /\mathcal{E}_\mathrm{EP} = 1,11,22,33,44,55$.  The peak  absorption  at  EPs (cyan solid) is  twice  larger  than   DPs (red dashed).  The $x$-axis denotes the frequency offset between the  field and the probed atomic transition.  
  }
  \label{fig:EPs_and_DPs}
\end{figure}
%----------------------------------------------------

When the pump amplitude  significantly  exceeds  $\mathcal{E}_\mathrm{EP}$ (at $\mathcal{E}\approx70\mathcal{E}_\mathrm{EP}$), pairs of resonances  merge again at ordinary degeneracies,  called diabolical points (DPs). The trajectories of the complex eigenvalues  of the Floquet Hamiltonian [see SI,~Eqs.~(C6-C7)]  are shown in \figref{EPs_and_DPs}(a). Panel (b) shows the absorption spectrum at varying pump intensities for the case where the pump couples  $2s^2\,(^1S)$ and  $2s2p\,  (^1P)$ [case $(ii)$ above]. The plot demonstrates  that  the peak of the ionization spectrum near the EPs is significantly larger than near the DPs (i.e.,  four-fold instead of two-fold), although the imaginary parts of the degenerate eigenvalues are  approximately equal.  This demonstrates the increased density of states at  EPs (similar to the effect shown in spontaneous emission near EPs in Maxwell's equations~\cite{pick2017general}).

In order to see  that the point at which the spectral peaks split is  an EP, we introduce a simplified model, where we keep only   three electronic states   (i.e., we keep only  the  ground state   and a pair of   AI state) and employ the rotating wave approximation~\cite{scully1999quantum} (RWA) to eliminate rapidly rotating terms due to  the pump and the probe. By using the RWA, a three-level system can be described  by the  stationary  effective Hamiltonian~\cite{scully1999quantum,kapralova2014helium}:
\begin{equation}
\mat{H} = 	\left(
	\begin{array}{c|cc}
	E_g+\hbar\omega_1  &  -\tfrac{\mu_{1}\mathcal{E}_1}{2} &  0  \\
		\hline
	-\tfrac{\mu_{1}\mathcal{E}_1}{2} & E_1-i\gamma_1   & -\tfrac{\mu_{2}\mathcal{E}_2}{2} \\
	0 &-\tfrac{\mu_{2}\mathcal{E}_2}{2} & E_2-i\gamma_2-\hbar\omega_2 
	\end{array}
	\right)
\label{eq:effective-H}
\end{equation}
Here,  $E_g$ is the real ground-state energy,  $E_{1,2}$ and $\gamma_{1,2}^{-1}$ are the excited-state energies and  lifetimes and 
$\mu_{1,2}$  are the complex transition  dipole moments of the  allowed transitions. 
$\omega_{1,2}$  and $\mathcal{E}_{1,2}$ are the frequencies and  amplitudes of the  probe and  pump fields respectively.
 The solid  lines in \eqref{effective-H} mark the  excited-state Hamiltonian, $\mat{H}_\mathrm{exc}$ (introduced in the introduction), whose  complex eigenvalues  ($\varepsilon_\pm$) and eigenvectors  ($\phi_\pm$) coalesce at an EP when the pump frequency and amplitude are~\cite{kapralova2014helium} 
\begin{equation}
\hbar\omega_2^\mathrm{EP} \equiv E_1 - E_2 +\frac{ \gamma_2-\gamma_1}{2}\frac{\mathrm{Im}\,\mu_2}{\mathrm{Re}\,\mu_2}
\quad,\quad
\mathcal{E}_2^\mathrm{EP} \equiv \pm\frac{ \gamma_1-\gamma_2}{2\,\mathrm{Re}\,\mu_2},
\end{equation}
More details are given in the SI. When the pump intensity is fixed at $\omega_2^\mathrm{EP}$ and its amplitude exceeds the critical value of $\mathcal{E}_2^\mathrm{EP}$, the ionization  peak splits into two  since the dressed-state energies, $\mathrm{Re}[\varepsilon_\pm]$, become non degenerate.  Similar to our conclusion here, it was also noted in~\citeasnoun{karapanagioti1995observation}  that  the  splitting  of spectral  lines occurs   when 
``the dipole coupling  is strong enough to compete with the autoionizing width.''  However, our  non-Hermitian approach  enables to identify this point   as an EP and, therefore,  opens the possibility of  exploring EP-related  effects near the splitting point.

To summarize, we presented an ~\emph{ab-initio} theory of  resonant photoionization. Our derivation produced accurate formulas for the ionization spectrum, which are   of current interest  due to recent developments in   experimental capabilities for  probing and controlling ionization processes.  By using NHQM, our theory avoids the need of computing the continuum states, which  are required by the traditional Feshbach formalism. As an application of our theory, we derive  a simple  expression  for the Fano asymmetry factor. Moreover, we  study autoionization in laser driven systems and show that  the splitting of spectral lines  occurs at EPs.   This    work  opens several  directions for future study.  For example, the  non-trivial topological phase associated with the EP~\cite{mailybaev2005geometric}  can be used to transition  between nearly degenerate  states in a topologically protected  manner~\cite{doppler2016dynamically}, and may have practical applications for controlling the ionization spectrum. 
Another example for an EP-related effect  is the enhanced density of states at the EP.
Previous work on spontaneous emission~\cite{pick2017general,Lin2016} shows that the  emission rate can be significantly enhanced by placing the emitter near pumped  resonators  with EPs (i.e., using systems that have optical gain).  Along similar lines, one can expect enhanced  autoionization rates in systems with gain of atomic population in particular AI states. Such a situation can be engineered, for example,   
(e.g., in systems with  cycling transitions~\cite{togan2011laser}). 
Finally, the present work treated   broadening of absorption lines due to   autoionization, but  can be extended  to include  other line-broadening mechanism (such as vibrational  and Doppler broadening) by combining NHQM with a Lindbladian formulation. 
\\

\begin{acknowledgments}
AP  is partially supported  by an Aly Kaufman Fellowship at the Technion. NM acknowledges the financial support of I-Core: The Israeli Excellence Center ``Circle of Light,'' and of the Israel Science Foundation Grant No. 1530/15. PRK acknowledges the financial support by the Czech Ministry of Education, Youth and Sports, program INTER-EXCELLENCE
 (Grant LTT17015). Finally,  the authors  thank Christiana Koch, Michael Rosenbluh,  Hossein Sadeghpour, Uri Peskin,  Saar Rahav, Gad Bahir, and  Ofer Neufeld for insightful discussions.
\end{acknowledgments}

\newpage
\section*{Supplementary Information}

%---------------------------------------------------------------
\section{Green's function Near  EPs}
\renewcommand{\theequation}{A\arabic{equation}}
\setcounter{equation}{0}
%---------------------------------------------------------------

Near an EP, the  non-Hermitian normal-mode  expansion formula for  the Green's function  [Eq.~(5)] breaks down. In this appendix, we
review the derivation of the modified expansion formula  which is valid at the EP, following \cite{pick2017general}.  Consider the parameter-dependent Hamiltonian:
\begin{gather}
H(\xi) = H_0 + \xi V
\end{gather}
where $H_0$ is  defective (i.e., the point $\xi = 0$ is an EP in parameter space). 
At the EP, the set of eigenvectors  is no longer a complete basis of the Hilbert space. To remedy this problem, we introduce additional Jordan vectors. At a second-order EP, the Jordan vector $\KET{j_\EP}$ is defined via the chain relations
\begin{gather}
H_0\KET{\phi_\EP} = \varepsilon_\EP\KET{\phi_\EP},\nonumber\\
H_0\KET{j_\EP} = \varepsilon_\EP\KET{j_\EP} + \KET{\phi_\EP}.
\end{gather}
where $\varepsilon_\EP$ and $\phi_\EP$ are the degenerate eigenvalue and eigenvector and  the self-orthogonality condition [$\BRACKET{\phi_\EP}{\phi_\EP} = 0$] is automatically satisfied. Following~\cite{Seyranian2003}, we choose the normalization conditions  $\BRACKET{\phi_\EP}{j_\EP} = \varepsilon_\EP^{-1}$ and  $\BRACKET{j_\EP}{j_\EP} = 0$.
Near the EP at $\xi = 0$, the Hamiltonian $H(\xi)$ has a pair of nearly degenerate eigenenergies and nearly parallel eigenvectors. They can be approximated by   a Puiseux series, which contains fractional powers in the small parameter $\xi$~\cite{Seyranian2003}:
\begin{gather}
\varepsilon_{{\pm}} = \varepsilon_\EP + \sqrt{\xi}\,\mathcal{V}+\mathcal{O}(\xi)\nonumber\\
\KET{\phi_\pm} =  \KET{\phi_\EP} + \sqrt{\xi}\,\mathcal{V}\KET{j_\EP}+\mathcal{O}(\xi)
\label{eq:defective-PT}
\end{gather}
where
\begin{gather}
\mathcal{V} = \sqrt{\frac{\BRAMKET{j_\EP}{V}{\phi_\EP}}{\BRACKET{j_\EP}{\phi_\EP}}}.
\end{gather}

Now, let us return to the modal expansion formula of $G$ [Eq.~(5)],  which is valid for $\xi\neq0$. Near the EP, the expansion is dominated by the two terms of the coalescing resonances.  Keeping just these two terms in the sum, we can write
\begin{gather}
G(\varepsilon) = \frac{1}{\varepsilon-\varepsilon_{+}}\frac{\KET{\phi_+}\BRA{\phi_+}}{({\phi_+}|{\phi_+})} +
\frac{1}{\varepsilon-\varepsilon_{-}}\frac{\KET{\phi_-}\BRA{\phi_-}}{(\phi_-|\phi_-)}
\label{eq:G_two_levels}
\end{gather}
Next, we  substitute the approximate expressions for $\KET{\phi_\pm}$ and $\varepsilon_\pm$ [\eqref{defective-PT}] into \eqref{G_two_levels} and,
by carefully taking the limit of $\xi\rightarrow0$, we obtain~\cite{pick2017general}
\begin{gather}
G(\varepsilon) = \frac{1}{(\varepsilon-\varepsilon_{\EP})^2}
\frac{\KET{\phi_\EP}\BRA{\phi_\EP}}{(\phi_\EP|j_\EP)} + 
\nonumber\\
\frac{1}{\varepsilon-\varepsilon_{\EP}}
\left(
\frac{\KET{\phi_\EP}\BRA{j_\EP}}{(\phi_\EP|j_\EP)} +
\frac{\KET{j_\EP}\BRA{\phi_\EP}}{(\phi_\EP|j_\EP)} 
\right)
\label{eq:G-at-EP}
\end{gather}
The double pole at $\varepsilon_\EP$ dominates the absorption spectrum near the EP.

%---------------------------------------------------------------
\section{Non-Hermitian Fano  factor}
\renewcommand{\theequation}{B\arabic{equation}}
\setcounter{equation}{0}
%---------------------------------------------------------------
In this appendix, we derive~Eq.~(8)  from the main text. The formula is obtained by comparing the ratio of the symmetric and antisymmetric parts of our new spectral formula [Eq.~(6)] and the  Fano lineshape near a single resonance [Eq.~(7)]. First, let us introduce the  dimensionless detuning parameter $x = \frac{\omega - \Omega}{\gamma/2}$ and rewrite~Eq.~(7) as 
\begin{equation}
S_F = 1 + \frac{2xq}{x^2+1} + \frac{q^2 - 1}{x^2 + 1}
\label{eq:fano explicit}
\end{equation}
Next, let us define the symmetric and antisymmetric parts of our absorption formula as
\begin{gather}
S_\mathrm{symm} = \frac{|\mathcal{E}|^2}{\hbar\pi}\frac{ \mathrm{Re}\,\mu_{if}^2 \mathrm{Im}\,\varepsilon_f }{(\omega - \mathrm{Re}\,\varepsilon_f)^2 + (\mathrm{Im}\,\varepsilon_f)^2}, \nonumber\\
S_\mathrm{asymm} = \frac{|\mathcal{E}|^2}{\hbar\pi}\frac{ \mathrm{Im}\,\mu_{if}^2 
( \mathrm{Re}\,\varepsilon_f - \omega)
}
{(\omega -  \mathrm{Re}\,\varepsilon_f)^2 + (\mathrm{Im}\varepsilon_f)^2},
\label{eq:fano schematic}
\end{gather}
where  introduced the shorthand notation $\mu_{if}^2 = \BRA{\phi_i^L}x \KET{\phi_f^R}\BRA{\phi_f^L}x \KET{\phi_i^R}$. 
By comparing \eqref{fano explicit} and \eqref{fano schematic} we find that
\begin{gather}
\frac{2q}{q^2 - 1} = 
\frac{ \mathrm{Im}\,\mu_{if}^2 }
{ \mathrm{Re}\,\mu_{if}^2 }.
\label{eq:roots}
\end{gather}
The solution of \eqref{roots} yields the Fano asymmetry factor [Eq.~(8)].  The sign of $q$ is determined by  the sign of $\mathrm{Im}\,\mu_{ij}^2$. When $q>0$, the absorption is stronger at frequencies higher than the  resonance frequency and weaker below the resonance. When $q<0$, the contrary is true.

 %---------------------------------------------------------------
\section{Absorption in laser-driven systems}
\renewcommand{\theequation}{C\arabic{equation}}
\setcounter{equation}{0}
%---------------------------------------------------------------

 %---------------------------------------------------------------
\subsection{The Floquet Hamiltonian}
%---------------------------------------------------------------

In this section, we explain how to construct the Floquet Hamiltonian and find its eigenvalues and eigenvectors, which appear in~Eq.~(10) in the main text. 
We wish  to solve the Floquet eigenvalue problem
\begin{equation}
\mathcal{H}\Phi_\alpha = \varepsilon_\alpha \Phi_\alpha,
\label{eq:floquet-eigen}
\end{equation}
where
\begin{equation}
\mathcal{H} \equiv H_0 + \mathcal{E}x\cos{\omega_0 t} - i\hbar\partial_t.
\label{eq:floquet-hamiltonian-app}
\end{equation}
In order to solve \eqref{floquet-eigen} numerically, 
we introduce $M$ temporal Fourier basis states, $f_m(t) = e^{i\omega mt}$, and $N$  spatial field-free states,  $\phi^\mathrm{FF}_\mu(x)$. Invoking  the completeness relation,
\begin{align}
\mathbb{1} = \sum_{m = 1}^M \KET{f_m(t)}\BRA{f_m(t)}
\otimes 
 \sum_{\mu = 1}^N\KET{\phi_\mu^\mathrm{R,FF}(x)}\BRA{\phi_\mu^\mathrm{L,FF}(x)},
\end{align}
we can rewrite \eqref{floquet-eigen} in matrix form:
\begin{equation}
\sum_{m,\nu} 
\BRA{n,\nu}\mathcal{H}
\KET{m,\mu}
\BRACKET{m,\mu}{\Phi_\alpha} = 
\varepsilon_\alpha
\BRACKET{n,\nu}{\Phi_\alpha} 
\label{eq:floquet-eigenproblem}
\end{equation}
or in shorthand notation: 
\begin{equation}
\overline{\overline{\mathcal{H}}}
\,\vec{\Phi}_\alpha = 
\varepsilon_\alpha \vec{\Phi}_\alpha,
\end{equation}
where  $\overline{\overline{\mathcal{H}}}$ is block diagonal, with block  size $M\times M$. 
The diagonal blocks are associated with the first and last terms in \eqref{floquet-hamiltonian-app}
\begin{equation}
\overline{\overline{\mathcal{H}}}_{\mu n,\nu n} = (\varepsilon_\mu^\mathrm{FF} + n\hbar\omega_0)\delta_{\mu,\nu}
\label{eq:H-floq-diag}
\end{equation}
and  the off-diagonal elements come from the second term:
\begin{equation}
\overline{\overline{\mathcal{H}}}_{\mu n,\nu n\pm1} = \tfrac{\mathcal{E}}{2}\BRA{\phi^\mathrm{L,FF}_\mu}x\KET{\phi^\mathrm{R,FF}_\nu}
\label{eq:H-floq-off-diag}
\end{equation}

%------------------------------------------------------
\subsection{Fermi-Floquet     absorption formula}
%------------------------------------------------------
In this appendix, we complete the derivation of~Eq.~(10) from the main text. Our derivation is inspired by Ref.~\cite{bilitewski2015scattering}, which analyzes scattering from a time-periodic potential. 
Consider an atom or molecule, which interacts with a laser at frequency $\omega_0$. The system is  described by the Hamiltonian $H_0$, whose eigenstates are  Floquet states, as explained  in the main text. The propagator of $H_0$ is defined via
\begin{gather}
i\hbar\partial_tU_0(t_0,t) = H_0(t)U_0(t_0,t). 
\label{eq:U-0}
\end{gather}
We also introduce  a weak laser, hereafter called ``the probe,'' with frequency $\omega$. The total Hamiltonian is 
\begin{gather}
H = H_0 + V,
\end{gather} 
where the interaction term is $V = \mathcal{E}xe^{i\omega t}$.
In order to derive Fermi-Floquet golden rule, we move to  the interaction picture, where  states and operators are defined as 
\begin{gather}
\KET{\Psi^I(t)} = U_0(t,t_0)\KET{\Psi(t)}\nonumber\\
\mathcal{O}^I(t) = U_0(t,t_0)\mathcal{O}U_0(t_0,t).
\end{gather}
Note that the total propagator, defined as
\begin{gather}
i\hbar\partial_tU(t_0,t) = H(t)U(t_0,t),
\label{eq:define-U}
\end{gather}
can be written as a product of the unperturbed and  interaction-picture propagators:
\begin{gather}
U(t_0,t) = U_0(t_0,t)U^I(t_0,t).
\label{eq:total-U}
\end{gather}
The last statement  can be verified by substituting  \eqref{total-U} into \eqref{define-U},  applying the chain rule to compute  $\partial_tU(t,t_0)$, and using \eqref{U-0} and $V^I = U_0(t,t_0)VU_0(t_0,t)$.

Next, we compute the transition amplitude between Floquet states $\KET{\Psi_f(t)}$ and $\KET{\Psi_i(t)}$, where $i$ and $f$ are super-indexes which denote the field-free state and the channel. The transition amplitude is
\begin{gather}
A(i\rightarrow f,t) = 
\BRA{\Psi_f(t)}U(0,t)\KET{\Psi_i(0)}=\nonumber\\
\BRA{\Psi_f(t)}U_0(0,t)U^I(0,t)\KET{\Psi_i(0)}=\nonumber\\
\BRA{\Psi_f(0)}U^I(0,t)\KET{\Psi_i(0)}
\label{eq:amplitude-calc}
\end{gather}
We use a Dyson series to express the interaction-picture propagator, $U^I$,in terms of $V^I$. Keeping terms up to the first order in $V^I$, one obtains:
\begin{equation}
U^I(t_0,t) = \mathbb{1} - \frac{i}{\hbar}\int_{t_0}^tdt'V^I(t') + \mathcal{O}(V^2).
\label{eq:Dyson}
\end{equation}
Substituting \eqref{Dyson} into \eqref{amplitude-calc}, one obtains
\begin{gather}
{A(i\rightarrow f,t) = 
\frac{-i}{\hbar}\int_0^t dt' \BRA{\Psi_f(0)}V^I(0,t') \KET{\Psi_i(0)} =} \nonumber\\
{\frac{-i}{\hbar}\int_0^t dt' \BRA{\Psi_f(0)}U_0(t',0)VU_0(0,t') \KET{\Psi_i(0)}  = }\nonumber\\
{\frac{-i}{\hbar}\int_0^t dt'
e^{-i(\varepsilon_i - \varepsilon_f)t'/\hbar}
 \BRA{\Phi_f(t')} V  \KET{\Phi_i(t')} . }
\end{gather}
Since the Floquet states $\Phi_{\alpha}(x,t)$ are periodic in time, one can decompose them into Fourier components
\begin{gather}
\Phi_{\alpha}(x,t) = \sum_n e^{i\omega_0 n t} \tilde{\phi}_{\alpha,n}(x)
\end{gather}
where the Fourier components of the wavefunction are $\tilde{\phi}_{\alpha,n}(x)\equiv \frac{1}{\sqrt{2\pi}}\int_{-\infty}^\infty dt\,\Phi_\alpha(x,t)e^{-i\omega_0 mt}$. Using this expansion, the transition amplitude becomes
\begin{align}
&{A(i\rightarrow f,t) = }
\nonumber\\
&{
\sum_{mn} 
\frac{-i\mathcal{E}}{\hbar}\int_0^t dt'
e^{-i(\varepsilon_i - \varepsilon_f-(n-m)\hbar\omega_0-\hbar\omega)t'/\hbar}
\BRA{\tilde{\phi}_{f,n}}x\KET{\tilde{\phi}_{i,m}}=}\nonumber\\
&{\mathcal{E}\sum_{mn}
\frac{e^{-i(\varepsilon_i - \varepsilon_f-m\hbar\omega_0-\hbar\omega)t/\hbar} - 1}{\varepsilon_i - \varepsilon_f-m\hbar\omega_0-\hbar\omega}
\BRA{\tilde{\phi}_{f,m+n}}x\KET{\tilde{\phi}_{i,m}}.}
\label{eq:step1}
\end{align}
The absorption spectrum can  be found by taking  the time average  of the transition probability:
\begin{gather}
{
S(\omega)\equiv \frac{1}{T} \int_0^T\!dt \,\frac{d}{dt}|A_{if}|^2 =  
2\mathrm{Re}\left[\frac{1}{T} \int_0^T\!dt \,A^*\frac{dA}{dt} \right],}
\label{eq:step2}
\end{gather}
where $T$ is a large integer multiple of the oscillation period $\frac{2\pi}{\omega_0}$.
Substituting \eqref{step1} into \eqref{step2} and neglecting rapidly oscillating terms, we obtain
\begin{gather}
\frac{1}{T}\int_0^T dt \,A^*\frac{dA}{dt}   = 
\frac{-i|\mathcal{E}|^2}{\hbar}
\sum_{mn\ell}
\frac{\BRA{\tilde{\phi}_{f,m+n}}x\KET{\tilde{\phi}_{i,m}}\BRA{\tilde{\phi}_{i,m}}x\KET{\tilde{\phi}_{f,m+\ell}}}
{(\varepsilon_i - \varepsilon_f-m\hbar\omega_0-\hbar\omega) }
\label{eq:last-step}
\end{gather}
Finally, we take the real part of \eqref{last-step} and arrive at 
\begin{gather}
{
S(\omega) = 
\frac{|\mathcal{E}|^2}{\hbar}
\mathrm{Im}
\left[\sum_{mn\ell}
\frac{\BRA{\tilde{\phi}^L_{f,m+n}}x\KET{\tilde{\phi}^R_{i,m}}
\BRA{\tilde{\phi}^L_{i,\ell}}x\KET{\tilde{\phi}^R_{f,\ell+m}}}
{\varepsilon_i - \varepsilon_f-m\hbar\omega_0-\hbar\omega}\right]}
\label{eq:floquet-app}
\end{gather}
The last formula can be rewritten compactly  as~Eq.~(10) in the main text.

%------------------------------------------------------
\subsection{Autoionization spectral formula   in  the  field-free basis}
%------------------------------------------------------

In order to evaluate our new formula for the absorption spectrum  [\eqref{floquet-app} or equivalently  Eq.~(10) from  the main text], we need to know the Fourier transforms of Floquet states. However, standard quantum chemistry methods  solve the field-free problem, and we would like to use the field-free basis states and avoid the formidable task of solving the Floquet eigenvalue problem for a multielectron atom or molecule. To this end, we expand the Fourier transforms of the Floquet states  in the basis of field-free  states. 
\begin{equation}
\KET{\tilde{\phi}_{\alpha,n}} = \sum_\mu 
\BRACKET{\phi^\mathrm{FF}_\mu}{\tilde{\phi}_{\alpha,n}}
\KET{\phi^\mathrm{FF}_\mu}
\label{eq:expand-ff}
\end{equation}
Substituting \eqref{expand-ff} into \eqref{floquet-app}, we obtain
\begin{gather}
{
S(\omega) = 
\frac{|\mathcal{E}|^2}{\hbar}
\mathrm{Im}
\left[\sum_{\substack{ mn\ell\\\mu\nu\sigma\tau}}
\BRACKET{\tilde{\phi}^L_{i,\ell}}{\phi^\mathrm{FF}_\tau}
\BRA{\phi^\mathrm{FF}_\tau}x\KET{\phi^\mathrm{FF}_\sigma}
\right.}
\nonumber\\
{
\left.
\frac{
\BRACKET{\phi^\mathrm{FF}_\sigma}{\tilde{\phi}^R_{f,\ell+m}}
\BRACKET{\tilde{\phi}^L_{f,n+m}}{\phi^\mathrm{FF}_\mu}}
{\varepsilon_i - \varepsilon_f-m\hbar\omega_0 - \hbar\omega}
\BRA{\phi^\mathrm{FF}_\mu}x\KET{\phi^\mathrm{FF}_\nu}
\BRACKET{\phi^\mathrm{FF}_\nu}{\tilde{\phi}^R_{i,n}}
\right]}
\label{eq:floquet-formula-in-ff}
\end{gather}
The transition dipole moments between field-free states, $\BRA{\phi^\mathrm{FF}_\tau}x\KET{\phi^\mathrm{FF}_\sigma}$, are obtained directly from 
available quantum chemistry codes.  By construction,  the expansion coefficients, $\BRACKET{\phi^\mathrm{FF}_\mu}{\tilde{\phi}^R_{\alpha,n}}$,
are the  components of the eigenvectors of the Floquet matrix:
\begin{equation}
 \BRACKET{\phi^\mathrm{L,FF}_\mu}{\tilde{\phi}^R_{\alpha,m}} = \BRACKET{m,\mu}{\Phi^R_\alpha}.
\end{equation}
The eigenvectors of the Floquet Hamiltonian, $\BRACKET{m,\mu}{\Phi^R_\alpha}$, are defined in \eqref{floquet-eigenproblem}.

%------------------------------------------------------
\subsection{EPs in the Floquet Hamiltonian}
%------------------------------------------------------

 To get an initial guess for the  location of the EP, it is convenient  project the full Hamiltonian onto the field-free excited states $\psi_2$ and $\psi_3$ and use the rotating wave approximation, which gives  the $2\times2$ Hamiltonian
\begin{equation}
H_\mathrm{exc} = 
\left( \begin{array}{cc}
E_2 & -\tfrac{\mu_{23}\mathcal{E}}{2} \\
-\tfrac{\mu_{32}\mathcal{E}}{2} & E_3 - \hbar\omega_0   \end{array} \right).
\end{equation}
Subtracting $E_2$ from the diagonal and introducing the definitions $\delta\equiv \mathrm{Re}[E_3 - E_2 - \hbar\omega_0]$ and $\Gamma \equiv -2\mathrm{Im}[E_3 - E_2]$, we obtain 
\begin{equation}
H_\mathrm{exc} = \left( {\begin{array}{cc}
  0&  -\tfrac{\mu_{23}\mathcal{E}}{2}  \\
-\tfrac{\mu_{23}\mathcal{E}}{2} &   \delta - i\tfrac{\Gamma}{2} 
  \end{array} } \right)  .
  \label{eq:reduced_Hamiltonian} 
\end{equation}
The characteristic polynomial of $H_\mathrm{exc}$ is
\begin{equation}
f(x) = x^2  -x(\delta - i\tfrac{\Gamma}{2}) -\tfrac{(\mu_{23}\mathcal{E})^2}{4} ,
\end{equation}
and EPs  occur when the discriminant of the polynomial vanishes:
\begin{equation}
\Delta^2 \equiv (\tfrac{\delta}{2} - i\tfrac{\Gamma}{4})^2 +\tfrac{(\mu_{23}\mathcal{E})^2}{4} = 0 
\end{equation}
Solving for $\mathcal{E}$ and $\delta$, we obtain the critical values~\cite{kapralova2014helium}:
\begin{equation}
\delta_\mathrm{EP} = \frac{\Gamma}{2}\frac{\mathrm{Im}\,\mu_{23}}{\mathrm{Re}\,\mu_{23}}
\hspace{1in}
\mathcal{E}_\mathrm{EP} = \pm\frac{\Gamma}{2\,\mathrm{Re}\,\mu_{23}}
\end{equation}
In the numerical calculation, we found  EPs in the large ($MN\times MN$)  Floquet Hamiltonian, including four field-free basis states and five Floquet bands, but found that the EP is obtained near the EP of the approximate $2\times2$ model. Specifically, we find an  EP of the full Floquet Hamiltonian at
$\delta \approx 1.001645\,\delta_\mathrm{EP}$ and $\mathcal{E}\approx 0.99420\,\mathcal{E}_\mathrm{EP}$.

 %---------------------------------------------------------------
\section{Electronic-structure of Helium}
\renewcommand{\theequation}{D\arabic{equation}}
\setcounter{equation}{0}
%---------------------------------------------------------------

\renewcommand\thefigure{\thesection.\arabic{figure}}    
\setcounter{figure}{0}  
\renewcommand{\thetable}{S\arabic{table}}
\setcounter{table}{0}

Table  S1  summarizes the complex energies and transition dipole moments for  lowest-energy single- and double-excitation states in helium.  The numerical values  were  obtained by Kapr{\'a}lov{\'a}-{\v{Z}}{\v{d}}{\'a}nsk{\'a}  \emph{et. al.}~\cite{kapralova2013excitation}, and are used in all the calculations in the main text. For copmarison, we show additional ab-initio results from~\citeasnoun{gilary2006ab}, which demonstrate good agreement between different methods for the set of energy levels that we consider.

%------------------------------------------------------
% FIGURE 1
%------------------------------------------------------
\begin{table}[!htb]
    \caption{Complex energies (top) and transition dipole moments (bottom) for helium, obtained from ab-initio calculations in~\citeasnouns{kapralova2013excitation,gilary2006ab}.}
    \begin{minipage}{.5\linewidth}
     % \caption{}
      \centering
         \begin{tabular}[b]{|c|c|c|c|c|c|}\hline
        &                &    Re[E]~\cite{gilary2006ab}         &    Re[E]~\cite{kapralova2013excitation}    & Im[E]~\cite{gilary2006ab}     		& Im[E]~\cite{kapralova2013excitation}       \\ \hline
      1& $1s^2$   & -2.90372     & -2.9035   &    0         		& 0      \\
      2& 1s2s       & -2.14597     & -2.1460     &    0         	&0     \\
      3& 1s2p       & -2.12384     & -2.1238     &  0            	& 0    \\
      4& $2s^2$  & -0.777868   & -0.7779     & 0.002271   	& -0.0023 \\
      5& $2p^2$  & -0.710500   & -0.7018      & 0.001181 	    	& -0.0012  \\
      6& 2s2p       & -0.693135  & -0.6930     &  0.0006865 	& -0.0007\\
      7& $2p^2$   & -0.621926  & -0.6216      & 0.000108 	& -0.0001\\ \hline
    \end{tabular}
    \label{table:complex_energies}
    \end{minipage}\\
    \begin{minipage}{.5\linewidth}
      \centering
       % \caption{}
           \begin{tabular}[b]{|c|c|c|c|}\hline
        			       & $\mu_R$~\cite{kapralova2013excitation} &    $\mu_I$~\cite{kapralova2013excitation}  &  $\lambda$ [nm]  \\ \hline
 $1\leftrightarrow3$&   0.4207		& 	0.00000189 &  58.44 (UV)   \\
$1\leftrightarrow6$&    0.03599	&    0.01299          & 20.61(UV) \\
$2\leftrightarrow3$&    2.9167		&   0.000004652  & 2058.47 (IR) \\
$2\leftrightarrow6$&   0.3130		&   -0.003598      &  31.36 (UV)\\
$3\leftrightarrow4$&   -0.1231 		&   -0.002554      & 33.85 (UV) \\
$3\leftrightarrow5$&    0.3288  	&   0.000193        & 32.04 (UV) \\
$3\leftrightarrow7$&    -0.1925 	&   0.0003475      & 30.33 (UV) \\
$4\leftrightarrow6$&   1.5227 		&   -0.00973         &536.95 (visible) \\
$5\leftrightarrow6$&   1.70545    	&  -0.003767        &5160.2 (IR)  \\
$6\leftrightarrow7$&    -2.1614 	&   -0.001007      & 638.15 (visible) \\ \hline
    \end{tabular}
        \vspace{-0.9in}
    \label{table:dipole_moments}
    \end{minipage} 
\end{table}
%------------------------------------------------------
\vspace{1in}

%\bibliography{absorptionEPjpcl}

\end{document}